\documentclass[letterpaper, 10 pt, conference]{ieeeconf}
\IEEEoverridecommandlockouts
\usepackage{graphics}
\usepackage{amsmath}
\usepackage{amssymb}
\usepackage[nocomma]{optidef}
\usepackage{xcolor}
\usepackage{siunitx}
\usepackage{flushend} 
\usepackage{tikz}
\usepackage{fontawesome5} 
\usetikzlibrary{fit, positioning}

\makeatletter
\let\NAT@parse\undefined
\makeatother
\usepackage{hyperref}

\mathtoolsset{showonlyrefs}

\newcommand{\prob}[1]{\mathcal{P}\left( #1 \right)}
\newcommand{\Real}[1]{\mathbb{R}^{#1}}
\newcommand{\NatSet}[1]{\mathbb{N}_{#1}}
\newcommand{\der}{\mathrm{d}}

\title{\LARGE \bf
Exploiting Prior Knowledge in Preferential Learning of Individualized Autonomous Vehicle Driving Styles
}

\author{Lukas Theiner$^{1}$, Sebastian Hirt$^{1}$, Alexander Steinke$^{1}$ and Rolf Findeisen$^{1}$
\thanks{The authors acknowledge Volkswagen AG for the fruitful collaboration and the resources and expertise provided.}
\thanks{$^{1}$Control and Cyber-Physical Systems Laboratory, Technical University of Darmstadt, Germany, {\tt\footnotesize lukas.theiner@iat.tu-darmstadt.de}\newline
        {\tt\footnotesize rolf.findeisen@iat.tu-darmstadt.de}}}%

\newcommand\copyrighttext{%
  \footnotesize \textcopyright \the\year{} IEEE. Personal use of this material is permitted. Permission from IEEE must be obtained for all other uses, including reprinting/republishing this material for advertising or promotional purposes, collecting new collected works for resale or redistribution to servers or lists, or reuse of any copyrighted component of this work in other works.}

\newcommand\copyrightnotice{%
\begin{tikzpicture}[remember picture,overlay]
\node[anchor=south,yshift=5pt] at (current page.south) {\fbox{\parbox{\dimexpr0.8\textwidth-\fboxsep-\fboxrule\relax}{\copyrighttext}}};
\end{tikzpicture}%
}

\begin{document}

\maketitle
\thispagestyle{empty}
\pagestyle{empty}
\copyrightnotice

\begin{abstract}
Trajectory planning for automated vehicles commonly employs optimization over a moving horizon -- Model Predictive Control -- where the cost function critically influences the resulting driving style. 
However, finding a suitable cost function that results in a driving style preferred by passengers remains an ongoing challenge.
We employ preferential Bayesian optimization to learn the cost function by iteratively querying a passenger's preference. 
Due to increasing dimensionality of the parameter space, preference learning approaches might struggle to find a suitable optimum with a limited number of experiments and expose the passenger to discomfort when exploring the parameter space. 
We address these challenges by incorporating prior knowledge into the preferential Bayesian optimization framework. 
Our method constructs a virtual decision maker from real-world human driving data to guide parameter sampling.
In a simulation experiment, we achieve faster convergence of the prior-knowledge-informed learning procedure compared to existing preferential Bayesian optimization approaches and reduce the number of inadequate driving styles sampled.
\end{abstract}

\section{Introduction}
In trajectory planning for automated driving, a multitude of requirements have to be considered, ranging from objective goals such as safety and time-optimality to highly subjective goals such as passenger comfort and perceived safety. 
\emph{Model predictive control} (MPC) \cite{rawlings2017model,findeisen2002introduction} has emerged as a suitable method for automated driving trajectory planning \cite{Liu2017_MPCforAD,bethge2020_MPCinAD}. 
MPC computes the vehicle trajectory by solving a series of optimal control problems subject to constraints.
While constraints are used to ensure that the resulting trajectory adheres to physical limitations and safety aspects, the formulation of the cost function has a strong influence on the resulting driving style of the automated vehicle. 
In addition to primary goals such as progressing along a desired route, the cost function is often used to consider secondary goals such as ride comfort or the mitigation of motion sickness \cite{steinke2022_traj_plan}.
However, selecting an appropriate cost function to align with user preferences remains a challenge in the context of automated vehicles and for control systems in general.

Recently, hierarchical learning and control frameworks have been explored that directly learn optimization parameters to achieve the desired closed-loop behavior.
These frameworks employ algorithms from reinforcement learning \cite{kordabad2023reinforcement,zanon2021safe} or Bayesian optimization \cite{paulson2023tutorial,hirt2024safe,hirt2024stability} at a higher level to adjust parameters of a lower-level MPC.

This hierarchical approach enables the incorporation of user preferences by taking advantage of algorithms from preference learning at the high level \cite{furnkranz2011preference}.
The application of preference learning in control systems has shown promising first results in learning controller parameters based on user feedback \cite{bemporad2021_preferencebasedMPCcalibration}.
Among the various preference learning methods, \emph{preferential Bayesian optimization} (PBO) \cite{wei2005_preference_gp, Brochu2007_active_preference_learning, lin2022_eubo} stands out for its ability to efficiently handle scenarios where evaluations are noisy and expensive \cite{cosner2022safety, shao2023preference}.
Such conditions are often encountered in applications like automated driving, where obtaining user feedback -- such as passenger evaluations of different trajectories -- can be costly and time-consuming, necessitating a sample-efficient learning procedure.

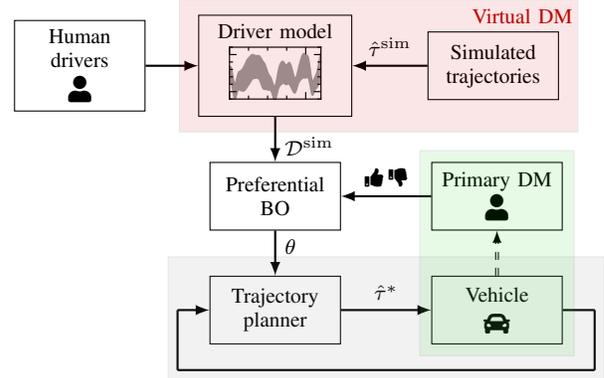
\begin{figure}
    \centering
    \newcommand{\myfigsize}{\footnotesize}
    \vspace*{0.4cm}
    \begin{tikzpicture}[
        block/.style={rectangle, draw, minimum width=1.7cm, minimum height=0.9cm},
        dashedblock/.style={rectangle, draw, minimum width=1.7cm, minimum height=0.9cm},
        groupblock1/.style={rectangle, draw, inner sep=0.2cm,fill=green!80!black,opacity=0.1},
        groupblock2/.style={rectangle, draw, inner sep=0.3cm,fill=red!80!black,opacity=0.1},
        groupblock3/.style={rectangle, draw, inner sep=0.2cm,fill=gray,opacity=0.1},
        arrow/.style={-latex, thick},
        line/.style={thick},
        dashedarrow/.style={-latex, double, dashed}
    ]
    \node[block] (pbo) {
        \parbox{1.5cm}{\centering\myfigsize Preferential \\ BO}
    };
    \node[dashedblock, above=0.6cm of pbo] (drivermodel) {
        \parbox{1.8cm}{\centering\myfigsize Driver model \\ \includegraphics[width=0.8\linewidth]{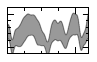}}
    };
    \node[dashedblock, right=1cm of drivermodel] (sim) {
        \parbox{1.5cm}{\centering\myfigsize Simulated trajectories}
    };
    \node[block, left=0.7cm of drivermodel] (humandriver) {
        \parbox{1.5cm}{\centering\myfigsize Human drivers \\ \normalsize \faUser}
    };
    \node[dashedblock, right=1.2cm of pbo] (primarydm) {
        \parbox{1.5cm}{\centering\myfigsize Primary DM \\ \normalsize \faUser}
    };
    \node[dashedblock, below=0.6cm of primarydm] (car) {
        \parbox{1.5cm}{\centering\myfigsize Vehicle \\ \normalsize \faCar}
    };
    \node[block, left=1.2cm of car] (tp) {
        \parbox{1.5cm}{\centering\myfigsize Trajectory planner}
    };
    \node[right=0.3cm of car.east] (fb1) {};
    \node[below=0.55cm of fb1] (fb2) {};
    \node[left=0.3cm of tp.west] (fb4) {};
    \node[below=0.55cm of fb4] (fb3) {};
    \draw[arrow] (humandriver) -- (drivermodel);
    \draw[arrow] (sim) -- node[midway, above] {\myfigsize $\hat{\mathbf{\tau}}^\mathrm{sim}$} (drivermodel);
    \draw[arrow] (drivermodel) -- node[pos=0.65, right] {\myfigsize $\mathcal{D}^\mathrm{sim}$} (pbo);
    \draw[arrow] (primarydm) -- node[midway, above] {\scriptsize \faThumbsUp\;\faThumbsDown} (pbo);
    \draw[dashedarrow] (car) -- (primarydm);
    \draw[arrow] (pbo) -- node[pos=0.35, right] {\myfigsize $\theta$} (tp);
    \draw[arrow] (tp) -- node[midway, above] {\myfigsize $\hat{\mathbf{\tau}}^*$} (car);
    \draw[line] (car.east) -- (fb1.center);
    \draw[line] (fb1.center) -- (fb2.center);
    \draw[line] (fb2.center) -- (fb3.center);
    \draw[line] (fb3.center) -- (fb4.center);
    \draw[arrow] (fb4.center) -- (tp.west);
    \node[groupblock1, fit=(primarydm)(car), inner ysep=0.15cm,inner xsep=0.15cm] {};
    \node[groupblock3, fit=(car)(tp)(fb1)(fb2)(fb3)(fb4), inner ysep=0.12cm,inner xsep=0.0cm, yshift=0.13cm] {};
    \node[groupblock2, fit=(drivermodel)(sim), inner ysep=0.2cm,inner xsep=0.25cm, label={[anchor=north east, xshift=2.7cm] \myfigsize\color{red!75!black} Virtual DM}] (virtualdm) {};
    \end{tikzpicture}
    \vspace*{-0.2cm}
    \caption{Preferential Bayesian optimization (PBO) provides parameters $\theta$ to the trajectory planner based on feedback by the primary decision-maker. The proposed prior-knowledge-informed PBO exploits data obtained from a virtual decision-maker, utilizing human driving data and simulations.}
    \vspace*{-0.5cm}
    \label{fig:overview}
\end{figure}

We focus on applying PBO to learn the parameters of an MPC-based trajectory planner such that the resulting system behavior aligns with individual user preferences.
Using passenger feedback in the form of pairwise preferences, our aim is to efficiently adjust the driving style of the automated vehicle to improve on subjective goals such as passenger comfort and perceived safety while satisfying objective physical and safety constraints.
Although PBO proved to be suitable for learning preference-based controller parameterizations, it becomes intractable for high-dimensional parameter spaces.
Specifically, we observed that an impractical number of closed-loop experiments is needed when tuning more than 3 parameters based on user preferences in our specific scenario.
Furthermore, exploring potential driving styles during the learning procedure can be associated with a high degree of discomfort for the passenger when sampling inferior parameterizations.

\newpage
These challenges motivate the main contribution of our work.
We incorporate prior knowledge about preferred driving styles in the form of a driver-model into the PBO procedure.
Specifically, we construct a probabilistic model of human driving from real-world data; see Fig. \ref{fig:overview}.
Our method aims to significantly reduce sample complexity and accelerate convergence to user-preferred driving styles. 
Additionally, our approach also reduces passenger discomfort during learning by focusing the exploration to sensible planner parameterizations according to the prior data.

The remainder of the article is structured as follows.
In Section \ref{sec:fundamentals}, we provide fundamentals about parameterized MPC-based trajectory planning and the associated learning problem and review preferential Bayesian optimization.
Section \ref{sec:methods} introduces our \emph{prior-knowledge-informed} PBO algorithm by detailing how we initialize it with data based on observations of real-world human drivers.
In a simulation study presented in Section \ref{sec:experiments}, we show the performance of the preference learning and planning framework. 

\section{Problem Formulation and Fundamentals}
\label{sec:fundamentals}
We consider an automated vehicle controlled by an optimization-based trajectory planner.
The planner's parameters are to be learned to achieve a desired driving style according to the passengers' preferences.
We first establish the MPC-based trajectory planning problem, before giving an introduction to preferential Gaussian processes and preferential Bayesian optimization. 

\subsection{Trajectory Planning for Automated Driving}
\newcommand{\predik}[2]{\hat{#1}_{#2|k}}
\newcommand{\predargs}{\predik{x}{i}, \predik{u}{i}, i{+}k}
\newcommand{\Nx}{n_\mathrm{x}}
\newcommand{\Nu}{n_\mathrm{u}}
\newcommand{\Np}{n_\mathrm{p}}
We consider automated vehicles with discrete dynamics
\begin{equation} \label{eq:model}
    x_{k+1} = f(x_k, u_k, k),
\end{equation}
with states $x_k \in \Real{\Nx}$, inputs $u_k \in \Real{\Nu}$, stage index $k \in \mathbb{N}$, and nonlinear dynamics $f: \Real{\Nx} \times \Real{\Nu} \times \mathbb{N} \rightarrow \Real{\Nx}$.

The purpose of the trajectory planner is to compute trajectories $\hat{\tau}^*_{k}$ in a receding-horizon manner, that allow the vehicle to progress safely and comfortably along a given road.
At stage $k$, the trajectory planner determines an optimal trajectory $\hat{\tau}^*_{k} = \left( \{\predik{x}{i}\}_{i\in\NatSet{N}}, \{\predik{u}{i}\}_{i\in\NatSet{N{-}1}} \right)$ over the current planning horizon of length $N\in\mathbb{N}$ by solving the parameterized optimal control problem
\begin{mini}||[2]<break>
{\mathbf{\hat{\tau}}_k}
{E_\theta(\predik{x}{N}) + \sum_{i=0}^{N-1} l_\theta(\predargs)}{\label{eq:ocp}}{}
\addConstraint{\forall i\in \NatSet{N-1}\colon}{\,}
\addConstraint{\hspace*{-1cm} \predik{x}{i+1}} {= \hat{f}(\predargs), \; \predik{x}{0}=x_k,}
\addConstraint{\hspace*{-1cm} 0} {\geq h(\predargs),\; 0 \geq h_N(\predik{x}{N}).}
\end{mini}
Here, $\NatSet{m}$ denotes the integer subset $\{0, 1, ..., m\}$.
At stage~$k$, a value predicted $i$ steps along the prediction horizon by the model $\hat{f}$ is denoted $\predik{\cdot}{i}$.
Constraints for ensuring safety and physical limits are given by $h(\cdot)$ and $h_N(\cdot)$, while $E_\theta(\cdot)$ and $l_\theta(\cdot)$ are the terminal and stage cost functions, depending on $\Np \in \mathbb{N}$ parameters $\theta \in \Theta \subset \mathbb{R}^{\Np}$.
The initial state for the current planning stage is denoted $x_k$.
In the following, our objective is to learn the parameters $\theta$ to achieve a preferred driving style.

\subsection{Preferential Gaussian Processes}
 \label{sec:fundamentals-preference-gp}
To optimize the parameters of \eqref{eq:ocp}, we employ preferential Bayesian optimization, which uses a surrogate model that captures the relationship between the parameters $\theta$ and the user preferences.
We construct this surrogate model by applying \emph{Gaussian process} (GP) preference learning to pairwise comparisons of parameterizations obtained from the user.

GP preference learning \cite{wei2005_preference_gp} allows the derivation of a probabilistic model of an unknown latent utility function $g~:~\mathbb{R}^{n_\xi}\to\mathbb{R}, \xi\mapsto g(\xi)$ from pairwise preference data.
A GP $\gamma(\xi) \sim \mathcal{GP}(m(\xi), k(\xi, \xi^\prime))$ is fully defined by its prior mean function $m: \mathbb{R}^{n_\xi} \to \mathbb{R}, \xi \mapsto \mathbb{E}[\gamma(\xi)]$, which represents the expected utility, and the prior covariance function $k: \mathbb{R}^{n_\xi} \times \mathbb{R}^{n_\xi} \to \mathbb{R}, (\xi, \xi') \mapsto \mathrm{Cov}[\gamma(\xi), \gamma(\xi')]$, which encodes the similarity between instances in terms of their utilities.

In a preference GP, the data consist of pairwise comparisons that express preferences.
Specifically, the training data set $\mathcal{D} = (\Xi, \mathcal{C})$ consists of $N_\mathrm{\Xi} \in \mathbb{N}$ inputs $\Xi = \{\xi_l\}_{l \in \mathbb{I}}$ with the index set $\mathbb{I} = \{1,\dots,N_\mathrm{\Xi}\}$ and $N_\mathcal{C} \in \mathbb{N}$ comparisons 
\begin{equation}
    \mathcal{C} = \{ (i,j)_l\}_{l \in \{1,...,N_\mathcal{C}\}},
\end{equation}
with ${i,j\in\mathbb{I}},\,{i\neq j}$.
If the tuple $(i,j)$ is contained in $\mathcal{C}$, input $\xi_i$ is preferred over $\xi_j$, i.e., the first index of the tuple refers to the preferred input and the second index to the less preferred.

\newcommand{\lat}{\mathbf{g}}
Before incorporating the observed preferences, a prior is placed over the latent utility function, which expresses the initial belief about the function's properties.
As in \cite{wei2005_preference_gp}, we assume a zero prior mean and a Gaussian prior covariance.
Thus, the prior for the latent utility at the training inputs is given by
$$
    \prob{\lat} = (2\pi)^{-N_\Xi/2} |\Sigma|^{-1/2} \exp ( -\tfrac{1}{2} \lat^\top \Sigma^{-1} \lat),
$$
where $\lat=[g(\xi_1),\dots,g(\xi_{N_\Xi})]^\top$ and $\Sigma \in \Real{N_\Xi \times N_\Xi}$ with elements $\Sigma_{ij}=k(\xi_i , \xi_j)$.
 
Given the data set $\mathcal{D}$, the objective is to infer the posterior distribution of the latent utility function values $g(\xi)$.
The likelihood function for each pairwise comparison is modeled using the probit likelihood given by
\begin{equation} \label{eq:probitlikelihood_pbo}
    \prob{\mathcal{D}|\lat} = \prod_{(i,j)\in \mathcal{C}} \mathrm{\Phi}\left( \frac{g(\xi_i)-g(\xi_j)}{\sqrt{2} \sigma} \right).
\end{equation}
Here, $\mathrm{\Phi}(\cdot)$ is the cumulative distribution function of the standard normal distribution.
The probit likelihood is chosen for its ability to model noisy pairwise comparisons. The parameter $\sigma$ controls the level of noise in the preferences, allowing the model to account for conflicting comparison data, which may arise due to inconsistency of the user evaluation or measurement error \cite{wei2005_preference_gp}.

Since the likelihood is non-Gaussian, the integrals required for posterior predictions cannot be analytically computed.
Thus, we perform Laplace approximation and model the posterior $\prob{\lat|\mathcal{D}}$ using a Gaussian distribution centered at the maximum of the posterior.
Therefore, we first obtain the \textit{maximum a posteriori} (MAP) estimate $\hat{\lat}$ by maximizing the posterior distribution $\prob{\lat | \mathcal{D}}$, yielding the convex~\cite{wei2005_preference_gp} optimization problem 
$
    \hat{\lat} 
    = \arg\max_\lat  \left\{ \prob{\lat | \mathcal{D}} \right\}
    = \arg\min_\lat \left\{ -\ln \prob{\lat} - \ln \prob{\mathcal{D}|\lat} \right\}.
$

Predictions of the latent utility value $g(\xi_*)$ at any previously unseen test input $\xi_*$ are then given by the Gaussian distribution $\mathcal{N}(\mu_*, \Sigma_*)$ with
\begin{align}
    \mu_* &= \mathbf k_*^\top \Sigma^{-1} \hat{\lat}
    \\
    \Sigma_* &= k(\xi_*,\xi_*) - \mathbf k_*^\top (\Sigma + \hat{\Lambda}^{-1})^{-1} \mathbf k_* .
\end{align}
Therein, $\mathbf k_* = [k(\xi_*,\xi_1),\dots , k(\xi_*,\xi_{N_\Xi})]^\top$ and $\hat{\Lambda}$ is the negative Hessian of the log-likelihood at the MAP estimate, see \cite{wei2005_preference_gp} for details. 
The hyperparameters of kernel $k$ and the likelihood \eqref{eq:probitlikelihood_pbo} are inferred by evidence maximization \cite{wei2005_preference_gp}.

\subsection{Preferential Bayesian Optimization}
 \label{sec:fundamentals-preference-bo}
\emph{Preferential Bayesian optimization} (PBO) is an iterative optimization strategy for problems, where the objective is to optimize an unknown utility function based on the preferences of a \emph{decision-maker} (DM).
Since the relationship between the input parameters $\theta$, the resulting outcomes, and the DM’s utility is complex and cannot be expressed analytically, PBO provides an effective approach to efficiently explore the parameter space \cite{lin2022_eubo}.

Using PBO, we solve the black-box optimization problem
\begin{equation}
    \label{eq:pbo_problem}
    \theta^* = \arg \max_{\theta \in \Theta} \left\{ G(\theta) \right\},
\end{equation}
where $G: \Real{n_\text{p}} \to \Real{}, \theta \mapsto G(\theta)$  is the unknown utility function defined by the DM's preference.
To this end, we employ a preferential Gaussian process surrogate $\hat G(\theta)$ to model the unknown function $G(\theta)$.
We learn the model iteratively by observing comparisons in a structured way and repeatedly update them with the newly gathered information.
Specifically, in each iteration $n \in \mathbb{N}$, the PBO procedure
\begin{enumerate}
    \item selects the next two parameter sets of interest ($\theta_a$,~$\theta_b$) and queries the DM by conducting closed-loop experiments using the MPC-based trajectory planner \eqref{eq:ocp} for each parameter set,
    \item updates the training data set with the newly observed data point
    \begin{alignat*}{4}
        \Xi^{n+1} &\leftarrow \Xi^n &&\cup \{ \theta_a, \theta_b \},
        \\
        \mathcal{C}_\theta^{n+1} &\leftarrow \mathcal{C}_\theta^n &&\cup \{c\},
    \end{alignat*}
    where $c=(a,b)$ if $\theta_a$ is preferred over $\theta_b$ and $c=(b,a)$ otherwise,
    \item updates the (posterior) preferential GP model based on $\mathcal{D}^{n+1} = (\Xi^{n+1}, \mathcal{C}^{n+1})$.
\end{enumerate}

To structure the sequential learning process, an \emph{acquisition function} $\alpha: \mathbb{R}^{n_{\text{p}}} \times \mathbb{R}^{n_{\text{p}}} \to \mathbb{R}, \quad (\theta_a,\theta_b) \mapsto \alpha(\theta_a, \theta_b; \mathcal{D}^n),$ is employed to guide the selection of new parameter set comparisons toward the optimal parameters $\theta^*$ as $n$ increases.
It leverages the surrogate model $\hat{G}$ of the unknown function $G$ to quantify the utility of a comparison of parameter sets $\theta_a$ and $\theta_b$ based on the data collected up to iteration $n$.
Utilizing the uncertainty information from the surrogate model, the learning procedure balances exploration and exploitation within the parameter space.
In our case, the next comparison of parameter sets is then selected using the \textit{expected utility of the best option} (EUBO) acquisition function given by \cite{lin2022_eubo}
\begin{equation}
    \label{eq:eubo_acquisition_function}
    \alpha_{\text{EUBO}}(\theta_a, \theta_b) = \mathbb{E} [ \max \{ \hat G(\theta_a), \hat G(\theta_b) \} ].
\end{equation}
Here, the expectation \( \mathbb{E} \) is taken with respect to the posterior distribution of $\hat G$.
Essentially, \eqref{eq:eubo_acquisition_function} optimizes the expected gain in utility when observing an additional comparison between $\theta_a$ and $\theta_b$ \cite{lin2022_eubo}.

\section{Proposed Method}
\label{sec:methods}
We employ the outlined PBO strategy to iteratively learn the parameters of a trajectory planner in an automated vehicle to adjust the driving style according to the passenger's preference. 
We are especially interested in an approach suitable for real-world experiments, where a human decision-maker iteratively adjusts the driving style through his preference feedback while being driven.

Although Bayesian optimization methods are generally sample efficient and thus suitable for optimization of expen\-sive-to-evaluate functions, learning a high number of parameters with standard PBO can be intractable.
We thus outline how we make expensive real-world experiments feasible by exploiting prior knowledge about preferred driving styles.
The proposed method queries a secondary, \emph{virtual decision-maker}, basing the decision on a probabilistic model of real-world human driving behavior.  

\subsection{Preferential BO with Multiple Decision Makers}
We consider two sources of information, the \emph{primary DM} -- for example, provided by real-world human-in-the-loop experiments -- and a \emph{virtual DM}. While the former represents the actual learning target, it is generally expensive to query.
The latter is, conversely, ``cheap'' to evaluate and even provides quantitative utility data $G^\mathrm{sim}(\theta)$, but might occasionally disagree with the primary DM due to a mismatch of its utility model and the unknown utility function of the primary DM. 
Hence, one still needs to be cautious when incorporating data that are not sourced by the primary DM.
In the proposed approach, we source data from a virtual DM, but mitigate the outlined challenges by the means of deliberate data selection and an adaption of the probit likelihood.

\subsection{Prior-knowledge-informed Preferential BO}\label{sec:method-priorinformedpbo}
Our approach of \emph{prior-knowledge-informed} PBO incorporates prior knowledge into the PBO learning procedure, as outlined in the following four steps:

\subsubsection{Human driver model}
We employ prior knowledge in the form of real-world human driving data, collected by measuring multiple trajectories of human driving along the same route.
The measured trajectories are preprocessed, spatially sampled, and assembled into a probabilistic, data-driven driver model by fitting a heteroscedastic Gaussian process.
Using a heteroscedastic GP, we allow for spatially varying noise; see \cite{kesting2007_mostlikeliheteroskedasticgp} for a detailed discussion.

\subsubsection{Virtual Decision Maker}
A virtual DM is created by defining a utility function $G^\mathrm{sim} : \Real{n_\mathrm{p}} \rightarrow \Real{}$, that assigns a utility value $G^\mathrm{sim}(\theta)$ to each possible parameter $\theta$ according to how well the driving style caused by $\theta$ fits with the human driver model.
The virtual DM always prefers the input with the highest utility $G^\mathrm{sim}(\theta)$.

Evaluating $G^\mathrm{sim}$ first computes a simulated trajectory $\mathbf{\tau}^\mathrm{sim}(\theta)$ of length $N$ by solving problem \eqref{eq:ocp}.
Note that the simulated trajectories can be pre-computed offline.
In addition, it evaluates the posterior of the human driver model GP at the same spatial sampling points as the trajectory, yielding a multivariate normal distribution $\mathbf{\tau}^\mathrm{human} \sim \mathcal{N}(\mu, \Sigma)$.
The utility is then defined as the log-likelihood of the planned trajectory under this distribution.

\subsubsection{Data Generation and Selection}
We utilize the virtual DM to generate a data set $\mathcal{D}^\mathrm{sim}$ of simulation queries $\Xi^\mathrm{sim}$ with associated comparisons $\mathcal{C}^\mathrm{sim}$ based on the virtual DM's preference. 
Therefore, we sample query points on a grid, covering the full parameter space, and compute the utilities $G^\mathrm{sim}(\theta)$, from which pairwise comparison data $\mathcal{C}^\mathrm{sim}$ is straightforwardly obtained.
These data are comprised of preference relations based on prior knowledge on human driving behaviors and are used to inform the PBO algorithm by initializing it with $\mathcal{D}^1 \leftarrow \mathcal{D}^\mathrm{sim}$.

An important step is the selection of query points and comparisons that should be carried over from the virtual DM to the PBO procedure. 
While the purpose of incorporating these data is to accelerate the PBO by reducing the amount of exploration needed, it is crucial to not precondition the PBO to only search in the region of the virtual DM's optimum.
We consistently achieved good results when choosing the pairs with the highest utility difference in simulation for the initialization of the data set for the PBO.

\subsubsection{Probit Likelihood}
In PBO, the noise parameter $\sigma$ in the probit likelihood \eqref{eq:probitlikelihood_pbo} allows for potentially conflicting preference relations and can be interpreted as Gaussian noise on the latent utility values.
As we want to trust the primary DM's preference relations if it disagrees with the virtual DM, we adjust the likelihood \eqref{eq:probitlikelihood_pbo} to be 
\begin{multline} \label{eq:probitlikelihood_pbo_modified}
    \prob{\mathcal{D},\mathcal{D}^\mathrm{sim}|\lat} = \prod_{(i,j)\in \mathcal{C}\setminus\mathcal{C}^\mathrm{sim}} \mathrm{\Phi}\left( \frac{g(\xi_i)-g(\xi_j)}{\sqrt{2} \sigma} \right)  
    \\
    \cdot\prod_{(i,j)\in \mathcal{C}^\mathrm{sim}} \mathrm{\Phi}\left( \frac{g(\xi_i)-g(\xi_j)}{\sqrt{2} \sigma^\mathrm{sim}} \right),
\end{multline}
with $ \sigma^\mathrm{sim} \gg \sigma$.
In practice, we choose $\sigma^\mathrm{sim} = \beta \cdot \sigma$, with $\beta$ being a design parameter and $\sigma$ being a hyperparameter of the preferential GP.
This way, we assume a higher uncertainty in the virtual DM's preferences, leading to higher trust in the primary DM.

\section{Evaluation}
\label{sec:experiments}
We first outline the experimental setup, including the formulation of the trajectory planning optimal control problem and the probabilistic human driver model.
Subsequently, we present and discuss the results of a simulation study evaluating and comparing prior-knowledge-informed PBO with standard PBO.

\subsection{Trajectory Planner}
We formulate the trajectory planner as a receding-horizon optimal control problem of the form \eqref{eq:ocp}, similarly to \cite{steinke2022_traj_plan}.
Under the common assumption that the desired comfortable driving styles do not approach the limits of handling, we choose a simple kinematic vehicle model.
The vehicle motion is modeled in a curvilinear coordinate system around a reference path with varying curvature $\kappa_\mathrm{ref}(s)$, resulting in the continuous motion model given by
\begin{equation} \label{eq:model_time}
    \dot{x}
    = \begin{bmatrix}
        \dot{v} \\ \dot{d} \\ \dot{\chi}
    \end{bmatrix}
    = f_{\mathrm c}(x, u, s)
    = \begin{bmatrix}
        a_\mathrm{x} \\ v \sin{(\chi)} \\ v \kappa - \dot s(x, s) \kappa_\mathrm{ref}(s) 
    \end{bmatrix}.
\end{equation}
Here, $s$ is the progress along the path, $d$ and $\chi$ are the signed lateral and angular deviations, $v$ is the vehicle velocity and the progress velocity is given by
\begin{equation}
    \dot s (x, s) = \frac{v \cos{(\chi)}}{1 {-} \kappa_\mathrm{ref}(s) d} .
\end{equation}
The longitudinal acceleration $a_\mathrm{x}$ and curvature $\kappa$ are the inputs $u = \left[ a_\mathrm{x},\; \kappa \right]^\top$. 
To simplify the consideration of the varying curvature of the reference path $\kappa_\mathrm{ref}(s)$, and other path-varying quantities, like the road widths, we transform the model into a spatial domain.
This is achieved by multiplying \eqref{eq:model_time} with the reciprocal of the progress velocity, resulting in the spatial model
\begin{equation} \label{eq:spatial}
   \der x / \der s = \dot{s}^{-1}(x, s) \cdot f_{\mathrm c}(x, u, s).
\end{equation}
This model is then discretized using the explicit Euler scheme with step lengths $h_k = s_{k+1}-s_k$ to obtain the discrete prediction model $\hat{f}(x,u,k)$.
We add the constraint
\begin{equation}
    \forall k \in \NatSet{N{-}1}: \quad 
    \left( \frac{a_{\mathrm{x},k}}{a_\mathrm{x,max}} \right)^2 
    + 
    \left( \frac{a_{\mathrm{y},k}}{a_\mathrm{y,max}} \right)^2 
    \leq 1
\end{equation}
with $a_{\mathrm{y},k}=v_k^2 \cdot \kappa_k$ and conservatively sized $a_\mathrm{x,max}$, $a_\mathrm{y,max}$
to stay within traction limits, and within the range where the fidelity of the kinematic model is sufficient.
The inputs are bounded according to the physical limitations of the vehicle, and the states are bounded by
$
\forall k\in\NatSet{N}: 
0 {\,<\,} v_{k} {\,\leq\,} v_\mathrm{max}, \;
\kappa_\mathrm{ref}(s_k){\,\cdot\,}d_k{\,<\,}1, \;
|\chi_k|{\,<\,}\frac{\pi}{2} \, ,
$
to avoid singularities in the model and
$
\forall k \in \NatSet{N}:
d_{\mathrm{min}}(s_k)\leq~d_k\leq~d_{\mathrm{max}}(s_k)$
to consider varying road bounds.
The stage cost of the optimal control problem \eqref{eq:ocp} is given by
\begin{equation} \label{eq:stage_cost}
    l_\theta(x_k,u_k,k) = \Delta t_k + w_\theta^\top \phi(x_k,u_k,k) \Delta t_k \, .
\end{equation}
It is comprised of the travel time $\Delta t_k =  \dot{s}^{-1}(x_k, s_k) \cdot h_k$ to  incentivize the vehicle to progress along the road,
and weighted features 
\begin{equation} \label{eq:features}
   \phi(x_k,u_k,k) = [ a_{\mathrm{x,pos},k}^2, a_{\mathrm{x,neg},k}^2, a_{\mathrm{y},k}^2, j_{\mathrm{x},k}^2, j_{\mathrm{y},k}^2 ]^\top
\end{equation}
for enhancing ride comfort.
As features in the following experiments, we consider squared longitudinal and lateral acceleration and jerk, the latter being approximated by forward differences.
Positive and negative longitudinal accelerations are considered separately to allow for different penalization of acceleration and braking.
Since the influence of the weights on the optimal trajectory tends to scale with the order of magnitude, we parameterize the weights by their exponents, i.e. $w_\theta^i = 10^{(\theta^i)}$.

The optimal control problem \eqref{eq:ocp} is solved using the interior-point solver \textsc{Ipopt} \cite{waechterbiegler2005_ipopt}.

\subsection{Human Driver Model} \label{sec:drivermodel-experiment}
\begin{figure}
    \centering
    \includegraphics{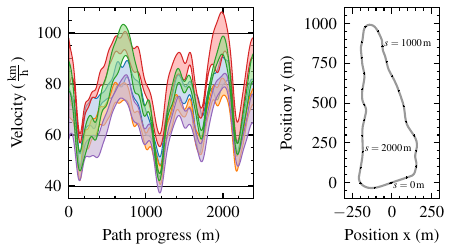}
    \caption{
        Left: Heteroscedastic Gaussian process models of five different observed human driving styles. The colored areas show the $2\sigma$ confidence regions. Right: Overview of the track layout. 
    }
    \vspace{-0.5cm}
    \label{fig:drivermodel}
\end{figure}

The probabilistic human driver model is constructed from real-world manual driving trajectories from a total of $22$ laps around a \SI{2.4}{\kilo\meter} long handling track.
We obtained data of five different driving styles from multiple advanced drivers being instructed to demonstrate either \textit{comfortable} or \textit{quick} driving.
Multiple laps for each driving style were recorded and sampled at spatial discretization points $s_k$.
The track layout and the velocity profiles of the resulting driver models are shown in Fig. \ref{fig:drivermodel}.

In the following, the driver model for the virtual DM in the prior-knowledge-informed PBO will be based on data from four of the five observed driving styles, while the fifth (green) will serve as the primary DM in a simulation experiment. 

We sample the parameter space at a grid with $5$ samples per dimension and compute the utility for each sample.
Of the ${>}4.8$ million possible pairs, we decide to pick the $3^{n_\mathrm{p}} = 243$ pairs with the greatest utility difference and add them to the dataset $\mathcal{D}^\mathrm{sim}$ for the prior-knowledge-informed PBO.

\subsection{Simulation Experiment}
To evaluate our method, we conduct an experiment with a simulated primary DM, which operates based on the same principles as the virtual DM, but with different data (see Section \ref{sec:drivermodel-experiment}). 
In each PBO iteration, the DM computes trajectories covering the full lap around the handling course for both queried parameter sets by solving the optimal control problem \eqref{eq:ocp}.
It further evaluates both trajectories under the human driver model to obtain the respective utility as outlined in Section \ref{sec:method-priorinformedpbo}, based on which the preference is selected.
We conducted multiple runs, using both our prior-knowledge-informed PBO procedure as well as a standard PBO procedure from literature. 

\begin{figure}
    \centering
    \includegraphics{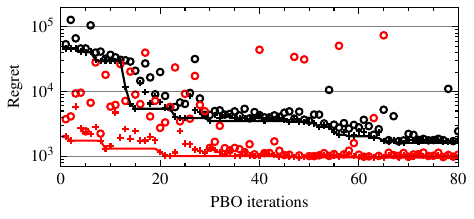}
    \caption{
    Regret of each queried sample over the course of the PBO iterations when using prior-knowledge-informed PBO (red) and a standard PBO approach (black). 
    Preferred samples are indicated by a $+$, less preferred samples are indicated by $\circ$.
    The continuous lines show the simple regret.
    }
    \vspace{-0.2cm}
    \label{fig:regret}
\end{figure}

Figure \ref{fig:regret} shows the regret for each queried sample during the course of a single PBO run, where regret is defined as the difference between the utility achieved by PBO at each iteration and the best possible utility \cite{garnett2023bayesian}.
While we employ information about the regret to gain insight into the proposed method, note that this information will not be available in a real-world experiment.
This is because the only feedback regarding the performance of the current query is the comparative judgment made by the primary DM.
The preferred sample in each comparison is marked with a $+$, while $\circ$ indicates the sample less preferred by the DM. 
The continuous lines show the simple regret, which is the regret of the best sample seen so far \cite{garnett2023bayesian}.

The proposed prior-knowledge-informed PBO approach is clearly able to accelerate convergence compared to standard PBO.
In particular, the standard PBO requires around $60$ iterations ($120$ parameter queries) to reach the regret score that the prior-knowledge-informed PBO already achieves in iteration $1$.
The standard PBO converges after $70$ iterations but is only able to learn an inferior parameterization.

\begin{figure}[t]
    \centering
    \includegraphics{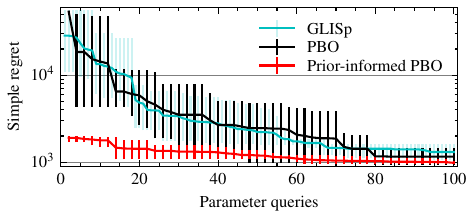}
    \caption{
    Comparison of simple regret over the course of the learning process for prior-knowledge-informed PBO (red), standard PBO (black), and GLISp \cite{bemporad2021_glisp,bemporad2021_preferencebasedMPCcalibration} (cyan), averaged over 5 trials each. The error bars show the full observed range.
    Note that GLISp only queries a single new parameter per iteration and compares with the best-so-far, while PBO queries a pair of parameter sets. Hence, with the same experimental budget we grant GLISp 100 iterations, while PBO only runs for 50 iterations.
    }
    \vspace{-0.5cm}
    \label{fig:regret_multi}
\end{figure}

Repeating the experiment for a total of $10$ trials, with half of the trials utilizing prior-knowledge-informed PBO, proves that our method consistently outperforms the standard PBO. The mean progression of simple regret in all trials is depicted in Figure \ref{fig:regret_multi}, with the error bars covering the range observed in all trials. For comparison, Figure \ref{fig:regret_multi} also shows the results using the method in \cite{bemporad2021_glisp, bemporad2021_preferencebasedMPCcalibration}, achieving a performance similar to that of standard PBO.

\begin{figure}
    \centering
    \includegraphics{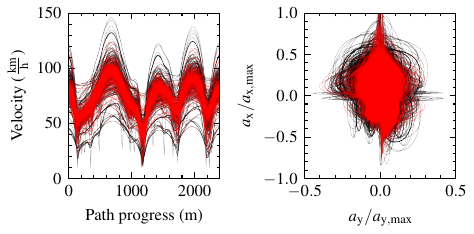}
    \caption{
        Comparison of queried driving styles over all PBO iterations when using the data-based human driver model as prior knowledge (red) and without prior knowledge (black), all 10 experiment trials aggregated. Left: velocity profile. Right: normalized lateral and longitudinal accelerations.
    }
    \vspace{-0.5cm}
    \label{fig:iterations}
\end{figure}

Furthermore, the prior-knowledge-informed PBO is able to focus exploration during learning to a region of reasonable driving styles. 
From the first iteration, it queries fewer samples with low utility scores than the standard PBO, with even the less preferred sample often outperforming both samples of the standard PBO (Fig. \ref{fig:regret}).
When comparing the velocity profiles of all queried driving styles (Fig. \ref{fig:iterations}, left), it becomes apparent that the prior-knowledge-informed PBO mostly avoids querying extreme driving styles that do not align with the human driver model used as prior knowledge, such as excessive speeds and unnecessary low cornering speeds. However, both occur during the standard PBO. 
Similarly, analyzing the accelerations acting on the passenger during the experiment (Fig. \ref{fig:iterations}, right), one can see that the regions of especially uncomfortable simultaneous longitudinal and lateral accelerations, which are usually avoided by drivers \cite{bosetti2014_human_acceleration}, are also not queried when using prior-knowledge-informed PBO.
Consequently, the proposed prior-knowledge-informed PBO approach for parameter learning not only lowers the overall required experiment time but also avoids querying of extreme driving styles, reducing discomfort acting on the passengers during learning. 

\begin{figure}[ht]
    \centering
    \includegraphics{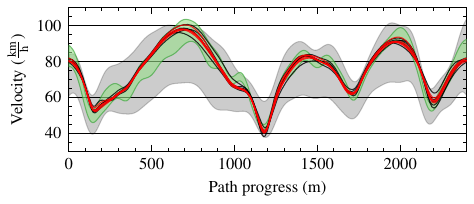}
    \caption{
        Trajectories resulting from parameters with largest latent utility after $50$ iterations of standard PBO (black) and prior-knowledge-informed PBO (red). The green and gray areas are the $2\sigma$ confidence regions of the simulated primary DM's driver model, and the virtual DM's driver model, respectively.
    }
    \vspace{-0.5cm}
    \label{fig:final_trajs}
\end{figure}

The trajectories resulting from the final parameters, obtained by maximizing the posterior mean of the latent utility function after $50$ iterations, are shown in Figure \ref{fig:final_trajs} and show a good fit to the driver model of the simulated primary DM.

\section{Conclusions}
We presented an approach to individualize the driving style of an automated vehicle based on feedback in the form of passenger preferences.
Specifically, this was done through online-learning the cost function of an MPC-based trajectory planner using preferential Bayesian optimization.
We proposed a strategy that allows PBO to exploit prior knowledge about preferred driving styles by introducing a secondary virtual decision maker.
The decisions of the virtual DM are based on a probabilistic model of human driving behavior constructed from real-world human driving data.
Thus, we achieved faster convergence of the PBO and a more pleasant driving experience during learning, which we demonstrated in a simulation experiment.

\bibliographystyle{IEEEtran}
\bibliography{literature}

\end{document}